\documentclass[aps,prd]{revtex4}

\usepackage{graphicx}
\usepackage{amsmath}
\usepackage{amssymb}

\def\lg{{\mathchoice{~\raise.58ex\hbox{$<$}\mkern-14.8mu\lower.52ex\hbox{$>$}~}
                    {~\raise.58ex\hbox{$<$}\mkern-14.8mu\lower.52ex\hbox{$>$}~}
                    {\raise.59ex\hbox{{$\scriptscriptstyle <$}}\mkern-12.8mu%
                     \lower.01ex\hbox{{$\scriptscriptstyle >$}}}   {}   }}
\def\gl{{\mathchoice{~\raise.58ex\hbox{$>$}\mkern-12.8mu\lower.52ex\hbox{$<$}~}
                    {~\raise.58ex\hbox{$>$}\mkern-12.8mu\lower.52ex\hbox{$<$}~}
                    {\raise.62ex\hbox{{$\scriptscriptstyle >$}}\mkern-12.0mu%
                     \lower.05ex\hbox{{$\scriptscriptstyle <$}}}  {}    }}

\def\ca{{\mathchoice{~\raise.58ex\hbox{$c$}\mkern-9.0mu\lower.52ex\hbox{$a$}~}
                    {~\raise.58ex\hbox{$c$}\mkern-9.0mu\lower.52ex\hbox{$a$}~}
                    {\raise.59ex\hbox{{$\scriptscriptstyle c$}}\mkern-7.0mu%
		     \lower.01ex\hbox{{$\scriptscriptstyle a$}}}   {} 	}} 
\def\ac{{\mathchoice{~\raise.58ex\hbox{$a$}\mkern-10.0mu\lower.52ex\hbox{$c$}~}
                    {~\raise.58ex\hbox{$a$}\mkern-10.0mu\lower.52ex\hbox{$c$}~}
		    {\raise.62ex\hbox{{$\scriptscriptstyle a$}}\mkern-9.0mu%
		     \lower.05ex\hbox{{$\scriptscriptstyle c$}}}  {} 	}}

\newcommand{\be}{\begin{equation}}
\newcommand{\ee}{\end{equation}}
\newcommand{\ba}{\begin{eqnarray}}
\newcommand{\ea}{\end{eqnarray}}
\newcommand{\ban}{\begin{eqnarray*}}
\newcommand{\ean}{\end{eqnarray*}}
\newcommand \nn {\nonumber}
\newcommand{\sla}{\!\!\!/ \,}

\begin{document}

\title{Ghosts in Keldysh-Schwinger Formalism}

\author{Alina Czajka}

\affiliation{Institute of Physics, Jan Kochanowski University, Kielce, Poland}

\author{Stanis\l aw Mr\' owczy\' nski}

\affiliation{Institute of Physics, Jan Kochanowski University, Kielce, Poland}
\affiliation{National Centre for Nuclear Research, Warsaw, Poland}

\date{March 14, 2014}

\begin{abstract}

We discuss how to introduce Faddeev-Popov ghosts to the Keldysh-Schwinger formalism describing equilibrium and non-equilibrium statistical systems of quantum fields such as the quark-gluon plasma which is considered. The plasma is assumed to be homogeneous in a coordinate space but the momentum distribution of plasma constituents is arbitrary. Using the technique of generating functional, we derive the Slavnov-Taylor identities and one of them expresses the ghost Green's function, which we look for, through the gluon one.  As an application, the Green's function of ghosts is used to compute the gluon polarization tensor in the hard loop approximation which appears to be automatically transverse, as required by the gauge invariance.

\end{abstract}



\maketitle

\section{Introduction}

In field theories obeying a gauge symmetry the number of fields exceeds the number of physical degrees of freedom. The unphysical degrees of freedom can be eliminated completely by a properly chosen gauge condition. However, such a condition usually breaks the Lorentz covariance of the theory and computations get complicated. To get rid of unphysical degrees of freedom in a manifestly Lorentz covariant way, one introduces the fictitious fields known as Faddeev-Popov ghosts which play a crucial role in nonAbelian field theories where unphysical degrees of freedom interact with physical ones. The ghosts naturally appear in the path integral formulation of quantum theory as a tricky representation of a Jacobian of gauge transformation. Then, the generating functional of Green's functions, which is obtained in an explicit form, determines the propagator of free ghost field. This is almost everything we need to include the ghosts in perturbative diagrammatic calculations, see {\it e.g.} \cite{Pokorski}. In statistical field theory, which is formulated in several ways, the situation is more complicated.

In the  Matsubara or imaginary time formalism, which applies to equilibrium systems, the ghosts are needed even in an Abelian theory \cite{Kapusta-Gale}. However, such non-interacting ghosts serve only to cancel unphysical degrees of freedom in the ideal gas contribution. In nonAbelian theories the ghosts are also included in the Feynman rules but the ghost propagator is obtained automatically when the explicit form of generating functional is computed \cite{Kapusta-Gale,lebellac}, provided the fermionic ghost fields obey the bosonic periodic boundary conditions, as argued in \cite{Bernard:1974bq}, see also \cite{Hata:1980yr}.

Sometimes a real time contour is included in the Matsubara approach and then one deals with the real time formalism of equilibrium systems which allows one to study time-dependent phenomena. The physical and unphysical degrees of freedom of gauge fields are usually treated on the same footing \cite{Landsman:1986uw,lebellac}. The Faddeev-Popov ghosts are thermalised with the bosonic distribution function. Within the alternative `frozen ghosts' approach the unphysical degrees of freedom and ghosts are kept at zero temperature that is their free Green's functions have no thermal contribution \cite{Landshoff:1992ne,Landshoff:1993ag}.

The problem of ghosts is least understood in the Keldysh-Schwinger formalism which provides a natural framework to study statistical systems out of equilibrium \cite{Calzetta-Hu,Chou:1987}. The formalism is obviously applicable to equilibrium systems as well. The main difficulty is that the generating functional cannot be computed in an explicit form even in noninteracting theory because of, in general, unknown density operator which enters the generating functional. Nevertheless the functional provides various relations among the Green's functions. To get free propagators, which are the basis of perturbative calculus, one solves the respective equations of motion. It should be noted here that free functions of the Keldysh-Schwinger formalism are much reacher than those of usual vacuum field theory. (We use the term {\it  vacuum field theory} to contrast it with the {\it statistical field theory}.) The Green's functions carry information not only about microscopic degrees of freedom of the system but about its statistical features as well. And it is unclear how to proceed with ghosts - whether these unphysical particles are constituents of the system of gauge fields or should be merely included in scattering matrix elements. 

The Faddeev-Popov ghosts result from the gauge freedom of a theory. Therefore, the gauge symmetry should determine completely a structure of ghost sector of the theory and we demonstrate here that in the Keldysh-Schwinger formalism this is also the case. For this purpose we derive the Slavnov-Taylor identities of quantum chromodynamics and show that one specific identity provides the ghost Green's function expressed through the gluon one. In this way the missing element of the diagrammatic computation scheme is found. An attempt to derive the  Slavnov-Taylor identities within the Keldysh-Schwinger formalism was undertaken in \cite{Okano:2001id} but, as explained at the end of Sec.~\ref{sec-ST-identities}, the result was rather unsatisfactory. 

The system of quarks and gluons under consideration is, in general, out of equilibrium but the system is assumed here to be translationally invariant. It is thus homogeneous (in coordinate space) but the momentum distribution is arbitrary. In particular, the system can be strongly anisotropic. The translational invariance greatly simplifies our analysis, as each two-point function depends on its two arguments only through their difference. When the assumption of homogeneity is relaxed, the analysis gets very complicated - the equations of motionf and Slavnov-Taylor identities are rather complex. One has to refer to the so-called gradient expansion to simplify them but it causes new difficulties. For this reason we focus here on the homogeneous systems. A much longer analysis of inhomogeneous ones will be presented elsewhere. 

The paper is organized as follows. In Sec.~\ref{sec-KS-formalism} we introduce the Keldysh-Schwinger formalism in the context of QCD and in the subsequent section the generating functional is written down. Sec.~\ref{sec-free-funcs} is devoted to the {\it free} Green's functions. We start with the equation of motion in a general covariant gauge showing that covariant gauges different than the Feynman one produce ill-defined expressions in the Keldysh-Schwinger approach. Therefore, we use the the Feynman gauge. Sec.~\ref{sec-ST-identities} presents a derivation of the Slavnov-Taylor identities. One of them expresses the free ghost Green's function through the gluon one. As an application of the developed method we compute the gluon polarization tensor at one loop level where, as well known, the ghost loop contributes.  The tensor found in the hard loop approximation is shown to be automatically transverse as required by the gauge invariance. In Sec.~\ref{sec-conclusions} we summarize our study, list the conclusions and we give an outlook. Some formulas, which are needed to perform calculations, are collected in Appendices.

Throughout the paper we use the natural system of units with $c= \hbar =1$; our choice of the signature of the metric tensor is $(+ - - -)$. Lorentz indices are denoted with $\mu, \, \nu = 0, \, 1, \, 2, \,3$. The color indices of fundamental representation of ${\rm SU}(N_c)$ gauge group are $i,\, j = 1, \, 2, \dots N_c$ and those of adjoint one $a,\, b = 1, \, 2, \dots N_c^2 -1$. The field operators in the operator formulation of quantum field theory are denoted in the same way as the classical fields in the path integral formulation. 

\section{Keldysh-Schwinger formalism}
\label{sec-KS-formalism}

\begin{figure}[t]
\centering
\includegraphics[scale=0.4]{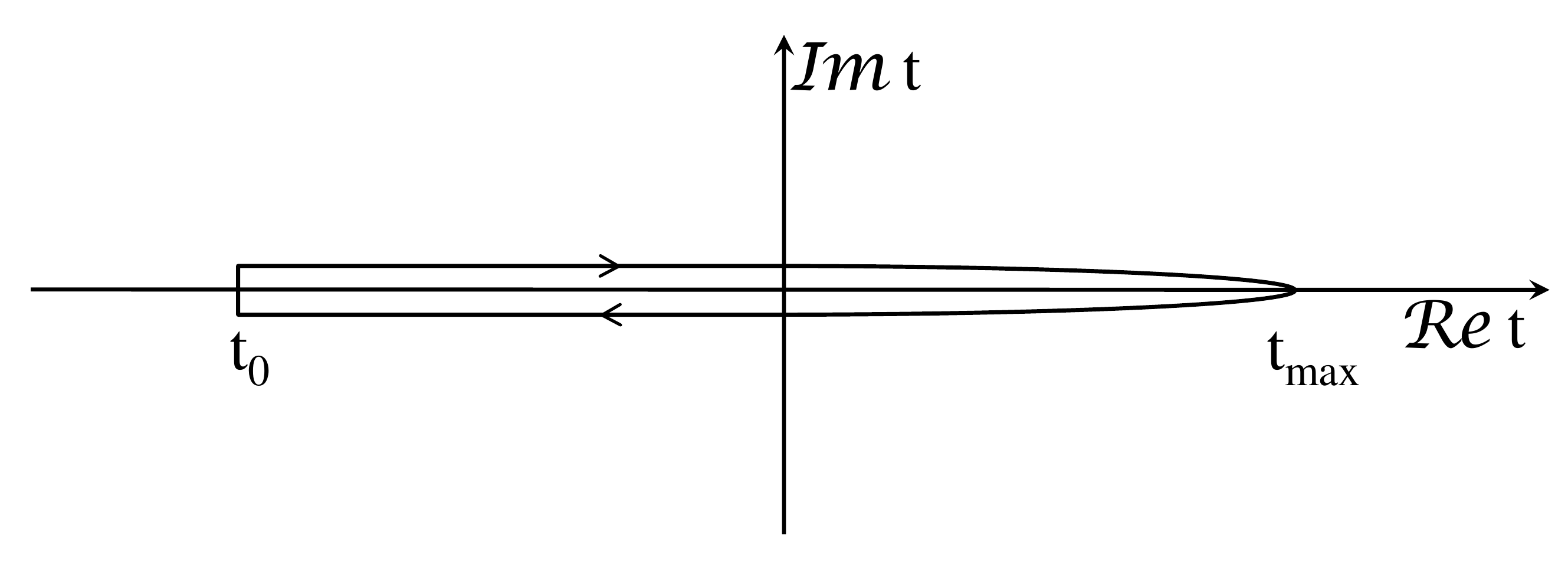}
\caption{The time contour of the Keldysh-Schwinger formalism.}
\label{fig-contour}
\end{figure}

We start our consideration with a brief presentation of the Keldysh-Schwinger formalism. Since the Yang-Mills fields are of our special interest, the formalism is presented in terms of Green's functions of the gauge vector field $A^a_\mu(x)$. The main object of the approach is the contour-ordered Green's function defined as
\be
\label{contour-GF}
i {\cal D}_{\mu\nu}^{ab}(x,y) \stackrel{{\rm def}}{=} \frac{{\rm Tr} \big[\rho (t_0) \, \tilde T A^a_\mu(x) A^b_\nu(y) \big]}{{\rm Tr}[\rho (t_0)]},
\ee
where the trace is understood as a summation over a complete set of states of the system ${\rm Tr} [\ldots]=\sum_\alpha <\alpha|\ldots|\alpha>$, $\rho (t_0)$ is a density operator at time $t_0$. The time arguments $x_0$ and $y_0$ are complex with an infinitesimal positive or negative imaginary part which locates them on the upper or lower branch of the contour shown in Fig.~\ref{fig-contour}. The real time $t_0$ is smaller than the real parts of $x_0$ and $y_0$ and the real time $t_{\rm max}$ is greater than the real parts of $x_0$ and $y_0$. The times $t_0$ and $t_{\rm max}$  are usually shifted to $- \infty$ and $+\infty$, respectively. The contour ordering operation $\tilde T$ is defined in the following way 
\be
\label{contour-time-ordering}
\tilde T A^a_\mu(x)A^b_\nu(y)  \stackrel{{\rm def}}{=} \Theta(x_0,y_0) A^a_\mu(x)A^b_\nu(y) + \Theta(y_0,x_0)A^b_\nu(y)A^a_\mu(x),
\ee
where $\Theta(x_0,y_0)$ is the contour step function defined as
\ba
\label{theta-function}
\Theta(x_0,y_0)=\left\{ \begin{array}{lll} 1,&& \textrm {if $x_0$ succeeds $y_0$ along the contour}, \\ 0, && \textrm {if
$y_0$ succeeds $x_0$ along the contour}. \end{array} \right.
\ea

The contour Green's function involves four Green's functions with real time arguments. They can be thought of as corresponding to propagation along the upper branch of the contour, along the lower one, from the lower branch to the upper one and from the upper branch to the lower one. This can be expressed in the following way
\ban
\label{real-time-GF}
{\cal D}_{\mu\nu}^{ab}(x,y) &=& \big({\cal D}_{\mu\nu}^{ab}\big)^c (x,y) \;\;\;\;\;\; \textrm{for $x_0,y_0$ on the upper branch}, 
\\[2mm]
{\cal D}_{\mu\nu}^{ab}(x,y) &=& \big({\cal D}_{\mu\nu}^{ab}\big)^a (x,y) \;\;\;\;\;\; \textrm{for $x_0,y_0$ on the lower branch} ,
\\[2mm]
{\cal D}_{\mu\nu}^{ab}(x,y) &=& \big({\cal D}_{\mu\nu}^{ab}\big)^{>} (x,y) \;\;\;\;\; \textrm{for $x_0$ on the lower branch, and $y_0$ on the upper one},
\\[2mm]
{\cal D}_{\mu\nu}^{ab}(x,y) &=& \big({\cal D}_{\mu\nu}^{ab}\big)^{<} (x,y) \;\;\;\;\; \textrm{for $x_0$ on the upper branch, and $y_0$ on the lower one}. 
\ean
The real-time argument Green's functions are thus defined as
\ba
\label{chronological-GF}
i \big({\cal D}_{\mu\nu}^{ab}\big)^c(x,y) & \stackrel{{\rm def}}{=} & 
\frac{{\rm Tr} \big[\rho(t_0) \, T^c A_\mu^a(x) A_\nu^b(y) \big]}{{\rm Tr}[\rho(t_0)]}, 
\\ [2mm]
\label{antichronological-GF}
i \big({\cal D}_{\mu\nu}^{ab}\big)^a(x,y) & \stackrel{{\rm def}}{=} & 
\frac{{\rm Tr} \big[\rho(t_0) \, T^a A_\mu^a(x) A_\nu^b(y) \big]}{{\rm Tr}[\rho(t_0)]}, 
\\ [2mm]
\label{bigger-GF}
i \big({\cal D}_{\mu\nu}^{ab}\big)^{>}(x,y) & \stackrel{{\rm def}}{=} & 
\frac{{\rm Tr} \big[\rho(t_0) \, A_\mu^a(x) A_\nu^b(y) \big]}{{\rm Tr}[\rho(t_0)]}, 
\\ [2mm]
\label{smaller-GF}
i \big({\cal D}_{\mu\nu}^{ab}\big)^{<}(x,y) & \stackrel{{\rm def}}{=} & 
\frac{{\rm Tr} \big[\rho(t_0) \, A_\mu^a(y) A_\nu^b(x) \big]}{{\rm Tr}[\rho(t_0)]},
\ea
where $T^c$ and $T^a$ are the usual chronological and antichronological time orderings. Directly from the definitions (\ref{chronological-GF}-\ref{smaller-GF}) one finds the following identities
\be
\label{rel-1}
{\cal D}^{c}(x,y)+{\cal D}^{a}(x,y) ={\cal D}^{>}(x,y) + {\cal D}^{<}(x,y),
\ee
\be
\label{rel-2}
{\cal D}^\ca (x,y)=\Theta (x_0-y_0) {\cal D}^\gl (x,y)+\Theta (y_0-x_0) {\cal D}^\lg (x,y),
\ee
which show that the four components of the contour Green's function are not independent from each other.

The contour Green's function carries information about microscopic interactions in the system under consideration and its statistical properties. The function ${\cal D}^c$ describes a particle disturbance propagating forward in time, and an antiparticle disturbance propagating backward in time. The meaning of ${\cal D}^a$ is analogous but particles are propagated backward in time and antiparticles forward. In the zero density limit ${\cal D}^c$ coincides with the usual Feynman propagator. The functions ${\cal D}^\lg$ play a role of the phase-space densities of (quasi-)particles, so they can be treated as quantum analogs of the classical distribution functions. Other Green's functions of gauge fields, which are used in Sec.~\ref{sec-applic}, are briefly discussed in Appendix~\ref{app-functions-gauge-KS}. Some formulas of the fermionic functions, which are needed to include quarks into our considerations, are collected in Appendix~\ref{app-functions-quark-KS}.

The main task of the Keldysh-Schwinger formalism is to derive contour Green's functions of the system under study. It can be achieved by solving properly approximated equations of motion analogous to the Dyson-Schwinger equation or by performing a perturbative expansion. Although the Green's functions of gauge fields are gauge dependent, they provide physical information which is independent of a gauge choice. For example, a spectrum of collective excitations obtained from the dispersion equation, where the polarization tensor enters, is gauge independent, provided the tensor is transverse. 

\section{Generating functional}
\label{sec-gen-fun}

The Keldysh-Schwinger approach can be formulated by defining the generating functional which is particularly useful to develop perturbative diagrammatic methods. We will need the functional to derive the Slavnov-Taylor identities discussed in Sec.~\ref{sec-ST-identities}. So, in this section we discuss the generating functional of quantum chromodynamics. To fix the notation and convention, which are used, we write down the fundamental Lagrangian of QCD as  
\be
\label{lagr-QCD}
\mathcal{L}_{\rm QCD} =-\frac{1}{4}F_a^{\mu\nu} F^a_{\mu\nu} 
+ {\bar \psi} ( i \gamma_\mu D^\mu - m ) \psi ,
\ee
where $F_a^{\mu\nu} \equiv \partial^\mu A^\nu_a - \partial^\nu A^\mu_a + g f^{abc} A^\mu_b A^\nu_c$ is  the strength tensor with $g$ being the QCD coupling constant and $f^{abc}$ the structure constants of ${\rm SU}(N_c)$ gauge group; $\psi$ is the quark field of mass $m$ and  $D^{\mu} \equiv \partial^\mu  - i g A^\mu_a \tau^a$ is the covariant derivative with $\tau^a$ being a generator of ${\rm SU}(N_c)$ group in the fundamental representation. The Lagrangian (\ref{lagr-QCD}) includes only one quark flavor but adding more flavors is straightforward.

Constructing the generating functional we follow \cite{Chou:1987} where the functional was given for non-gauge fields. So, the procedure has to be modified. The fundamental Lagrangian (\ref{lagr-QCD}) is replaced by the effective one 
\be
\label{lagr-gluons-eff}
\mathcal{L}_{\rm eff} = \mathcal{L}_{\rm QCD} - \frac{1}{2\alpha} \big(\partial^\mu A_\mu^a \big)^2 
- c^*_a(\partial^\mu \partial_\mu \delta^{ab} - g\partial^\mu f^{abc} A_\mu^c)c_b 
+ J^\mu_a A^a_\mu + \chi^*_a c_a  + \chi_a c^*_a.
\ee
The term, which follows $ \mathcal{L}_{\rm QCD}$, fixes the general covariant gauge and the subsequent one with $c^*$ and $c$ being the ghost Grassmann fields allows one to properly count the volume of a gauge orbit \cite{Pokorski}. The remaining three terms describe interactions of the fields $A, \, c$ and $c^*$ with external sources $J, \, \chi^*$ and $\chi$. The sources of ghosts are Grassmannian. The terms of interaction of quark fields with external sources are missing in Eq.~(\ref{lagr-gluons-eff}). Since we are mostly interested in the gauge fields, the quarks are ignored all together from now on to simplify the form of generating functional which is anyway rather complex.

Let us first write down the generating functional 
\ba
\label{W-KS-YM-0}
W_0[J, \chi, \chi^*] &=& N_0 \int_{\begin{subarray}{l} A(-\infty +i0^+,{\bf x})=A'({\bf x}) \\ 
A(-\infty -i0^+,{\bf x})=A''({\bf x})\end{subarray} } \mathcal{D}A(x) 
\int_{\begin{subarray}{l} c(-\infty +i0^+,{\bf x})=c'({\bf x}) \\ 
c(-\infty -i0^+,{\bf x})=c''({\bf x})\end{subarray} } \mathcal{D}c(x) 
\int_{\begin{subarray}{l} c^{*}(-\infty +i0^+,{\bf x})={c^{*}}'({\bf x}) \\ 
c^{*}(-\infty -i0^+,{\bf x})={c^{*}}''({\bf x})\end{subarray} } \mathcal{D}c^{*}(x) 
\nn \\ [2mm] && 
~~~~~~~~~~~~~~~~~~~~~~~~~~~~~~~~~~~~~~~~~~~~~~~~~~~~~~~~~~~~~~~~~~~~~~~~~~~~~~~~~~
\times \exp{\bigg[i\int_C d^4x \,\mathcal{L}_{\rm eff}(x) \bigg]} ,
\ea
which is labeled with the index `0' as it strongly resembles that of vacuum field theory. $N_0$ is the normalization constant and $\mathcal{D}A(x)$, $\mathcal{D}c(x)$, $\mathcal{D}c^{*}(x)$ are the standard functional integration measures of the fields $A(x) ,\, c(x) ,\, c^{*}(x)$ which depend on $x_0$ and ${\bf x}$ with $x_0$ from the contour $C$ shown in Fig.~\ref{fig-contour}. The fields obey the indicated boundary conditions at $t = -\infty \pm i0^+$ with the fields $A'({\bf x})$, $A''({\bf x})$, $c'({\bf x})$, $c''({\bf x})$,  ${c^{*}}'({\bf x})$, ${c^{*}}''({\bf x})$  which are now unspecified.  The integration over $x_0$ is performed along the time contour and we have denoted
\be
\label{denote}
\int_C d^4x \, \dots \equiv \int_C dt \int d^3x \, \cdots .
\ee
The functional $W_0[J, \chi, \chi^*]$ depends functionally on the boundary fields  $A'({\bf x})$, $A''({\bf x})$, $c'({\bf x})$, $c''({\bf x})$,  ${c^{*}}'({\bf x})$, ${c^{*}}''({\bf x})$ which are not shown as arguments to simplify the notation. If the boundary fields all vanish and the contour $C$ is replaced by the straight line from $-\infty$ to $\infty$, the functional (\ref{W-KS-YM-0}) coincides with the standard one of the vacuum field theory \cite{Pokorski}. 

The generating functional of Keldysh-Schwinger formalism is obtained from the functional (\ref{W-KS-YM-0}) by integrating it over the boundary fields $A'({\bf x})$, $A''({\bf x})$, $c'({\bf x})$, $c''({\bf x})$,  ${c^{*}}'({\bf x})$, ${c^{*}}''({\bf x})$ weighted with the density matrix 
\be
\label{dens-matrix}
\rho\big[A'({\bf x}),c'({\bf x}),{c^{*}}'({\bf x}) \big| A''({\bf x}),c''({\bf x}),{c^{*}}'' ({\bf x})\big] ,
\ee
which describes the system of fields at $t= -\infty$. The matrix is not really physical because of the unphysical degrees of freedom of gauge fields and of the ghosts which enter the formula (\ref{dens-matrix}). However, our results do not depend on a form of the density matrix. The complete generating functional equals
\ba
\label{W-KS-YM}
W[J, \chi, \chi^*]&=&N \int  DA'({\bf x}) \, DA''({\bf x})\, Dc'({\bf x}) \,Dc''({\bf x})\,D{c^{*}}'({\bf x}) \,D{c^{*}}''({\bf x})  
\nn \\ && \times ~~~~~~~~
\rho\big[A'({\bf x}),c'({\bf x}),{c^{*}}'({\bf x}) \big| A''({\bf x}),c''({\bf x}),{c^{*}}'' ({\bf x})\big] \;
W_0[J, \chi, \chi^*] .
\ea
The constant $N$ is chosen in such a way that $W[J=0, \chi=0, \chi^*=0]=1$.

It should be stressed that our results presented in the subsequent sections are fully independent of a form of the density operator which enters the generating functional (\ref{W-KS-YM}). So, we do not need to specify the operator but we could consider various forms of it. In particular, we could choose the boundary conditions of the ghost fields as $c'({\bf x}) = {c^{*}}'({\bf x}) = c''({\bf x}) = {c^{*}}'' ({\bf x}) = 0$ and the density operator, which acts on the ghost fields, as $|0\rangle \langle 0|$. Then, the ghost fields are treated exactly as in the vacuum theory. Consequently, the functional integral over the ghost fields, which is Gaussian, can be taken explicitly and one obtains the Fadeev-Popov determinant in the standard form. However, we do not follow this path.

The generating functional (\ref{W-KS-YM}) provides various Green's functions by differentiating it with respect to the sources $J$, $\chi$ or $\chi^*$. In particular, the two-point gluon contour function, which will be needed further on, is given as
\ba
\label{D-W-def}
i {\cal D}_{\mu\nu}^{ab}(x,y) &=& (-i)^2 \frac{\delta^2}{\delta J^\mu_a (x) \, \delta J^\nu_b (y)} 
W[J, \chi, \chi^*] \bigg|_{J=\chi=\chi^*=0}.
\ea 
Locating $x_0$ and $y_0$ on the upper or lower branch of the contour $C$, one gets the function ${\cal D}^c$, ${\cal D}^a$, ${\cal D}^>$ or ${\cal D}^<$. 

The functional (\ref{W-KS-YM}) can be used to derive the perturbative series which expresses the interacting Green's function ${\cal D}$ through the free Green's functions. The functions of free gluons $D$ are  found by solving the respective equations of motion, as it is done in the subsequent section, but there is a problem - also explained in the next section - with the free ghost functions which enter the perturbative expansion. Consequently, the expansion is not meaningful yet. 

Contrary to the vacuum field theory, the generating functional of the Keldysh-Schwinger formalism cannot be expressed in a closed explicit form even for a free theory because of the unspecified density operator which is present in Eq.~(\ref{W-KS-YM}). Nevertheless, the functional (\ref{W-KS-YM}) provides various relations among the Green's functions. In particular, one derives the Slavnov-Taylor identities which result from the gauge symmetry of the theory. The relations - generalizing the Ward-Takahashi identities of QED to Yang-Mills theories - are discussed in Sec.~\ref{sec-ST-identities}.

\section{Free Green's functions}
\label{sec-free-funcs}

In this section we derive an explicit form of the contour two-point Green's functions of free gauge fields. The free function is denoted by $D$ to distinguish it from the interacting one ${\cal D}$. A method  of derivation, which uses the equation of motion, is rather standard and it can be found, for example, in \cite{Mrowczynski:1992hq}.  Nevertheless there are some peculiarities because of the general covariant gauge we start with.  

The equation of motion of the contour Green's function of the free gluon field in a general covariant gauge reads
\ba
\label{eom-prop-A-1-arb}
\Big[\square_x g^{\mu\nu} -\Big(1-\frac{1}{\alpha}\Big) \partial_x^\mu \partial_x^\nu \Big] D_{\nu\rho}^{ab}(x,y) 
&=& g^\mu_{\;\rho} \delta^{ab} \delta_C^{(4)}(x,y),
\ea
where the contour Dirac delta function $\delta_C^{(4)}(x,y)$ is defined as
\ba
\label{delta-contour}
\delta^{(4)}_C (x,y)=\left \{ \begin{array}{llll}
\delta^{(4)}(x-y) & \qquad \textrm{for} & x_0,y_0& \textrm{from the upper branch},\\
0 & \qquad \textrm{for} & x_0,y_0& \textrm{from the different branches},\\
-\delta^{(4)}(x-y) & \qquad \textrm{for} & x_0,y_0& \textrm{from the lower branch}.
\end{array} \right.
\ea
As already mentioned, the system under consideration is homogeneous but the momentum distribution is, in principle, arbitrary. Due to the translational invariance, the propagators depend on the coordinates $x$ and $y$ only through their difference, that is $D(x,y) = D(x-y)$. 

One observes that using covariant gauges different than the Feynman one with $\alpha = 1$ leads to ill-defined expressions in the Keldysh-Schwigner formalism. The reason is the following. Performing the Fourier transformation of Eq.~(\ref{eom-prop-A-1-arb}), one finds that the structure of Lorentz indices of gluon Green's functions is 
\be
g^{\mu \nu} - (1-\alpha ) \frac{p^\mu p^\nu}{p^2} .
\ee
The contour Green's function includes the medium part describing gluons on the mass-shell $p^2 =0$ and consequently there appears a contribution to the Green's function proportional to 
\be
\delta(p^2)\Big( g^{\mu \nu} - (1-\alpha ) \frac{p^\mu p^\nu}{p^2} \Big) ,
\ee
where the second term is ill defined. It might well be that the term can be regulated by replacing it by a function with a double pole at $p^2=0$. However, such a prescription needs to be checked in detail. Instead, we simply get rid of the ill-defined term by choosing the Feynman gauge with $\alpha = 1$.

The Fourier transformed equation of motion of the Green's functions $D^\lg$ reads
\ba
\label{eom-FGG-4-em}
p^2 \big(D^{ab}_{\mu\nu}\big)^\lg (p) = 0.
\ea
The solutions can be written down as
\ba
\label{eom-FGG-deriv-1-em}
i \big(D^{ab}_{\mu\nu}\big)^> (p) &= &2\pi g_{\mu\nu} \delta^{ab} \delta(p^2)h(p), \\
\label{eom-FGL-deriv-1-em}
i \big(D^{ab}_{\mu\nu}\big)^< (p) &= & 2\pi g_{\mu\nu} \delta^{ab} \delta(p^2)g(p),
\ea
where $h(p)$ and $g(p)$ are unknown functions. Splitting the functions into positive and negative parts and using the fact that the difference $D^> - D^<$ must equal the Jordan function
\ba
\label{Green-Jordan-wigner5-em}
i \Big[\big(D^{ab}_{\mu\nu}\big)^>(p) - \big(D^{ab}_{\mu\nu}\big)^<(p)\Big] = 
- \frac{\pi}{E_p} g_{\mu\nu} \delta^{ab} \big[\delta(p_0-E_p) - \delta(p_0+E_p)\big],
\ea
one finds the unordered functions as
\ba
\label{D->}
\big(D^{ab}_{\mu\nu}\big)^>(p) &=& \frac{i\pi}{E_p} g_{\mu\nu} \delta^{ab} 
\Big[\delta(p_0 - E_p)\big(n_g({\bf p})+1\big) + \delta(p_0 + E_p)n_g(-{\bf p})\Big], \\
\label{D-<}
\big(D^{ab}_{\mu\nu}\big)^<(p) &=& \frac{i\pi}{E_p} g_{\mu\nu} \delta^{ab} 
\Big[\delta(p_0 - E_p)n_g({\bf p}) + \delta(p_0 + E_p)\big(n_g(-{\bf p})+1\big)\Big],
\ea
where $n_g({\bf p})$ is a distribution function of gluons which are assumed to be unpolarized with respect to spin and color degrees of freedom. The function is normalized in such a way that the gluon density is given as
\be
\rho_g = 2 (N_c^2 -1)  \int \frac{d^3p}{(2\pi)^3}\, n_g ({\bf p}) ,
\ee
where the factor of 2 takes into account two gluon spin states. So, the function $n_g({\bf p})$  takes into account only {\it physical} transverse gluons.

The Feynman $D^c$ and antiFeynman $D^a$ propagators obey the equation of motion
\ba
\label{eom-FaF}
p^2 \big(D^{ab}_{\mu\nu}\big)^\ca (p) &=& \mp \delta^{ab} g_{\mu \nu} ,
\ea
where the upper sign is for $c$ and the lower one for $a$. One finds the functions recalling the relation (\ref{rel-2}) which gives
\ba
\label{D-c}
\big(D^{ab}_{\mu\nu}\big)^c(p) &=&-g_{\mu\nu} \delta^{ab} \Big[ \frac{1}{p^2+i0^+} 
-\frac{i\pi}{E_p}\Big(\delta(p_0-E_p) n_g({\bf p}) + \delta(p_0+E_p) n_g(-{\bf p}) \Big)\Big]
\\[2mm]
\label{D-a}
\big(D^{ab}_{\mu\nu}\big)^a(p)&=& g_{\mu\nu} \delta^{ab} \Big[ \frac{1}{p^2-i0^+} 
+\frac{i\pi}{E_p}\Big(\delta(p_0-E_p) n_g({\bf p}) + \delta(p_0+E_p) n_g(-{\bf p}) \Big)\Big].
\ea
As seen, the functions  $D^c$ and $D^a$ contain the propagator parts combined with the medium contributions which vanish in the vacuum limit $n_g({\bf p}) \rightarrow 0$.  Then, we have usual propagators. 

The free Green's functions of a fermion field can be derived in a similar way by solving the appropriate equations of motion, see {\it e.g.} \cite{Mrowczynski:1992hq}. We do not derive the functions of quarks but in the Appendix~\ref{app-functions-quark-KS} we list some formulas which will be used in Sec.~\ref{sec-applic}. One could also find the Green's functions of ghost fields solving the equations of motion but it is fairly unclear what is the distribution function of ghosts. The Slavnov-Taylor identity, which is derived in the next section, allows one to resolve the ambiguity.

\section{Slavnov-Taylor identities}
\label{sec-ST-identities}

In this section we derive the Slavnov-Taylor identities of gluodynamics in the Keldysh-Schwinger approach. As already mentioned, we ignore quarks to simplify our considerations which are focused on the gauge and ghost fields. We do not refer to the BRST symmetry, which is used nowadays to obtain the Slavnov-Taylor identities, see {\it e.g.} \cite{Pokorski}, but we adapt the original Slavnov's method \cite{Slavnov:1972fg}, see also \cite{Faddeev-Slavnov}, to the Keldysh-Schwinger formalism. The point is that the BRST symmetry is {\it global} and then the fields, which are arguments of the density matrix present in the generating functional (\ref{W-KS-YM}), change under the BRST symmetry but the transformation properties of the matrix are unknown. To avoid the problem, we look how the generating functional (\ref{W-KS-YM}) changes under the {\it local} gauge transformation which, as suggested in \cite{Chou:1987},  vanishes at $t=-\infty$. In this way we first derive a general  Slavnov-Taylor identity and then we look for a specific relation which allows one to express the ghost Green's function through the gluon one. 

\subsection{Derivation of the general identity}
\label{subsec-general-identity}

To derive the  Slavnov-Taylor identities we first rewrite the functional (\ref{W-KS-YM}) in the form which strongly resembles that of vacuum field theory that is 
\be
\label{W-gluon-sti-sec}
W [J, \chi, \chi^*]  = N \int_{\rm BC} \mathcal{D}A \,\Delta[A]
\exp \Big[ i \int_C d^4 x \, \mathcal{L} \Big] ,
\ee
where we use a very compact notation 
\ba
\nn
\int_{\rm BC} \mathcal{D}A  \dots &\equiv &  
\int  DA'({\bf x}) \, DA''({\bf x})\, Dc'({\bf x}) \,Dc''({\bf x})\,D{c^{*}}'({\bf x}) \,D{c^{*}}''({\bf x}) \;
\label{sti-notation} 
\\[2mm] 
&& \times 
\rho\big[A'({\bf x}),c'({\bf x}),{c^{*}}'({\bf x}) \big| A''({\bf x}),c''({\bf x}),{c^{*}}'' ({\bf x})\big] 
\int_{\begin{subarray}{l} A(-\infty +i0^+,{\bf x})=A'({\bf x}) \\ 
A(-\infty -i0^+,{\bf x})=A''({\bf x})\end{subarray} } \mathcal{D}A(x) \dots
\ea
and 
\ba
\nn
\Delta[A] &\equiv &  
\int_{\begin{subarray}{l} c(-\infty +i0^+,{\bf x})=c'({\bf x}) \\ 
c(-\infty -i0^+,{\bf x})=c''({\bf x})\end{subarray} } \mathcal{D}c(x) 
\int_{\begin{subarray}{l} c^{*}(-\infty +i0^+,{\bf x})={c^{*}}'({\bf x}) \\ 
c^{*}(-\infty -i0^+,{\bf x})={c^{*}}''({\bf x})\end{subarray} } \mathcal{D}c^{*}(x)  
\\[2mm]
\label{sti-notation-delta}
&& ~~~~~~~~~~~~~~~~
\times \exp \Big[ -i \int_C d^4 x \Big(c^*_a(\partial^\mu \partial_\mu \delta^{ab} -g f^{abc} A_\mu^c \partial^\mu)c_b -   \chi^*_a c_a  -   \chi_a c^*_a\Big) \Big] , 
\ea
which is the analog of the Faddeev-Popov determinant. The Lagrangian in Eq.~(\ref{W-gluon-sti-sec}) is given by 
\be
\label{lagr-YM}
\mathcal{L} = \mathcal{L}_{\rm QCD} - \frac{1}{2} \big(\partial^\mu A_\mu^a \big)^2 
+ J^\mu_a A^a_\mu.
\ee
As already mentioned we use the Feynman gauge with $\alpha=1$.

The general Slavnov-Taylor identity results from the invariance of the generating functional (\ref{W-gluon-sti-sec}) with respect to
the infinitesimal gauge transformations  
\be
\label{gauge-trans-sti}
A_\mu^a \rightarrow (A_\mu^a)^U=A_\mu^a+f^{abc}\omega^b A_\mu^c 
-\frac{1}{g}\partial_\mu \omega^a +{\cal O}(\omega^2)
\ee
where the parameter $\omega$ is small, $|\omega| \ll 1$. We assume that the gauge transformation (\ref{gauge-trans-sti}) does not work  at $t=-\infty$, that is $\omega(t=-\infty, {\bf x})=0$, and consequently the density matrix $\rho$ in the expression (\ref{sti-notation}) remains unchanged.

Expressing the generating functional of gluodynamics (\ref{W-gluon-sti-sec}) by the transformed fields, one finds
\ba
\label{W-gauge-sti-1}
W'[J,\chi^*,\chi] &=& N \int_{\rm BC} \mathcal{D}A \, \Delta[A] 
\exp \Big\{i\int_C d^4x  \Big[\mathcal{L}  
- \frac{1}{g}M^{ab}\omega^b \partial^\nu A_\nu^a 
-\frac{1}{g} J^\mu_a \partial_\mu \omega^a
+J^\mu_a f^{abc} A_\mu^c \omega^b \Big]\Big\},
\ea
where the operator $M$, which functionally depends on $A^\mu_a$, equals 
\be
\label{M-def}
M_{ab}[A|x] \equiv -\partial^\mu\partial_\mu \delta^{ab} + g f^{abc}\partial^\mu A_\mu^c(x) .
\ee
We have also taken into account in Eq.~(\ref{W-gauge-sti-1}) that the QCD Lagrangian (\ref{lagr-QCD}) and the integration measure $\mathcal{D}A\,\Delta[A]$ are  invariant under the transformation (\ref{gauge-trans-sti}). 

The invariance of the theory with respect to the gauge transformation (\ref{gauge-trans-sti}) is reflected by the independence of the functional (\ref{W-gauge-sti-1}) of the parameter $\omega$. Therefore, the derivative of $W'[J,\chi^*,\chi] $ with respect to $\omega$ should vanish. However, if the functional (\ref{W-gauge-sti-1}) is independent of $\omega$, it is also independent of any function of $\omega$. For the reasons which will be clear later on, we are going to differentiate the functional (\ref{W-gauge-sti-1}) over the function $\xi_a(x)$ which is
\be
\label{f-ksi}
\xi_a(x)=M_{ab}[A|x] \,\omega_b(x) .
\ee
Introducing the operator $M^{-1}$, which is inverse to $M$ that is
\be
\label{eq-FG-M}
M_{ab}[A|x] \, M^{-1}_{bc}[A|x,y] =\delta_{ac} \delta_C^{(4)}(x,y),
\ee
one expresses the gauge parameter $\omega$ as
\be
\label{omega-f}
\omega_a(x)=\int_C d^4y \, M^{-1}_{ab}[A|x,y] \, \xi_b(y).
\ee
One guesses that $M^{-1}_{ab}[A|x,y]$ is related to the ghost Green's function, see below. Substituting the expression (\ref{omega-f}) into the functional (\ref{W-gauge-sti-1}), one finds
\ba
\nn
W'[J,\chi^*,\chi] &=& N \int_{\rm BC} \mathcal{D}A \,\Delta[A] \exp \Big\{i\int_C d^4x  \Big[\mathcal{L}(x)
- \frac{1}{g} \partial^\mu_{(x)} A_\mu^a(x) M_{ab}[A|x] \int_C d^4y \, M^{-1}_{bd}[A|x,y] \, \xi_d(y)   
\\ [2mm]
\label{W-gauge-sti-2}
&&~~
-\frac{1}{g} J^\mu_a(x) \partial_\mu^{(x)}   \int_C d^4y \, M^{-1}_{ad}[A|x,y] \, \xi_d(y)
+J^\mu_a(x) \, f^{abc} A_\mu^c(x) \int_C d^4y \, M^{-1}_{bd}[A|x,y] \, \xi_d(y) \Big]\Big\}.
\ea

The transformation (\ref{gauge-trans-sti}) can be treated as a change of integration variables but such a change cannot change a value of the integral. Thus, we get the condition
\be
\label{condition-gauge-sti}
\frac{\delta W'[J,\chi^*,\chi]}{\delta \xi_d(z)}\bigg|_{\xi=0} = 0 .
\ee
Differentiating the functional (\ref{W-gauge-sti-2}) with respect to $\xi$ and putting $\xi = 0$, one finds
\ba
\nn
&&\int_{\rm BC} \mathcal{D}A \, \Delta(A) \exp \Big[i \int_C d^4x\, \mathcal{L}(x) \Big] 
\\[2mm]
\label{STI-general-0} 
&& ~~~~~~\times \Big\{
- \partial_{(z)}^\mu A_\mu^d(z)  
- \int_C d^4 x \, J^\mu_a(x)\Big( \partial_\mu^{(x)}\delta^{ab} 
- g f^{abc} A_\mu^c(x) \Big) M^{-1}_{bd}[A|x,z]  \Big\} = 0 .
\ea
Performing the functional differentiation one should remember that $x_0$, $y_0$ and $z_0$ are on the contour and thus 
\be
\frac{\delta \xi_a(x)}{\delta \xi_b(y)} = \delta_{ab}\delta_C^{(4)}(x,y) .
\ee

Replacing the field $A_\mu^a(x)$ by the corresponding derivative 
\be
\label{replacement-sti}
A_\mu^a(x) \rightarrow \frac{1}{i}\frac{\delta}{\delta J^\mu_a(x)},
\ee
the relation (\ref{STI-general-0}) can be rewritten as 
\ba
\label{STI-general}
\bigg\{
i \partial_{(z)}^\mu \frac{\delta}{\delta J^{\mu}_d(z)} 
- \int_C d^4 x \, J^\mu_a(x) \bigg( \partial_\mu^{(x)} \delta^{ab}  
+ ig f^{abc} \frac{\delta }{\delta J^\mu_c(x)} \bigg)
M^{-1}_{bd}\Big[\frac{1}{i}\frac{\delta}{\delta J}\Big|x,z\Big] 
\bigg\} W[J,\chi^*,\chi]= 0,
\ea
which is the generalized Slavnov-Taylor identity in the Feynman gauge. In the subsequent section we discuss one specific identity following from Eq.~(\ref{STI-general}).

\subsection{The Slavnov identity for the gluon propagator}
\label{subsec-identity-gluon}

We are going to derive the identity which relates the gluon Green's function to the ghost one. Differentiating the general relation (\ref{STI-general}) with respect to $J^\nu_e (y)$ and putting $\chi=\chi^*=J=0$, we obtain
\ba
\label{STI-long-G-0}
\bigg\{
i \partial_{(z)}^\mu \frac{\delta^2}{\delta J^{\mu}_d(z) \, \delta J^{\nu}_e(y)} 
- \bigg( \partial_\nu^{(y)} \delta^{eb}  
+ i g f^{ebc} \frac{\delta }{\delta J^\mu_c(y)} \bigg)
M^{-1}_{bd}\Big[\frac{1}{i}\frac{\delta}{\delta J}\Big|y,z\Big] 
\bigg\} W[J,\chi^*,\chi] \bigg|_{\chi=\chi^*=J=0}= 0,
\ea
which requires further manipulations. Using the equation (\ref{eq-FG-M}) together with (\ref{M-def}), one observes that
\be
\label{eq-66}
\Big( - \partial^{(y)}_\nu \delta^{eb}  
+ g f^{ebc}  A_\nu^c(y) \Big)  
M^{-1}_{bd} [A|y,z] = - \partial^{(y)}_\nu \Delta_{ed} (y,z)  ,
\ee
where $\Delta_{ed} (y,z)$ is the Green's function of {\it free} ghost field obeying the equation of motion
\be
 - \partial_{(y)}^\nu \partial^{(y)}_\nu \Delta_{ed} (y,z) = \delta_{ed}\delta_C^{(4)}(y,z)  .
\ee
The equality (\ref{eq-66}) holds up to the function independent of $y$ which is eliminated due to the boundary conditions obeyed by $M^{-1} [A|y,z]$ and $\Delta (y,z)$.

Eq.~(\ref{eq-66}) allows one to write down the relation (\ref{STI-long-G-0}) as 
\ba
\label{STI-long-G}
\partial_{(z)}^\mu  \mathcal{D}_{\mu\nu}^{ab} (z,y)
= \partial^{(y)}_\nu \Delta_{ab} (y,z)  ,
\ea
where we have used the definition of the gluon Green's function (\ref{D-W-def}).  Eq.~(\ref{STI-long-G}) relates to each other the contour Green's functions of gluons and ghosts. Locating the time arguments $y_0$ and $z_0$ on the upper or lower branch of the contour shown in Fig.~\ref{fig-contour} we get the relations for the Green's functions of real arguments
\ba
\label{STI-long-G-1}
\partial_{(z)}^\mu  \big(\mathcal{D}_{\mu\nu}^{ab}\big)^\lg (z,y)
&=& \partial^{(y)}_\nu \big(\Delta_{ab}\big)^\gl (y,z)  ,
\\[2mm]
\label{STI-long-G-2}
\partial_{(z)}^\mu  \big(\mathcal{D}_{\mu\nu}^{ab}\big)^\ca (z,y)
&=& \partial^{(y)}_\nu \big(\Delta_{ab}\big)^\ca (y,z)  .
\ea
Since the system under study is translationally invariant, the Fourier transformed identity (\ref{STI-long-G}) gets the desired form
\ba
\label{gluon-prop-id}
- p^\mu \mathcal{D}_{\mu\nu}^{ab}(p) = p_\nu \Delta_{ab}(-p),
\ea
which relates the longitudinal part of the gluon Green's function to the free ghost function. Eq.~(\ref{gluon-prop-id}) also expresses the well-known fact that the longitudinal part of the gluon Green's function is not modified by interaction and consequently the polarization tensor, which results from the interaction, is purely transversal. 

As already mentioned, an attempt to derive the  Slavnov-Taylor identities within the Keldysh-Schwinger formalism was undertaken in \cite{Okano:2001id}. However, there were serious flaws in the derivation. The fields present in the generating functional (\ref{W-KS-YM}) were stated to obey periodic boundary conditions which effectively meant that the density matrix was diagonal. There was no justification for such an assumption. Since the global BRST transformation was used, the density matrix was assumed invariant under the transformation to guarantee the invariance of the generating functional. Again there was no justification for this assumption. It was also overlooked that the ghost contour Green's function includes the medium contribution, see the subsequent section, and consequently the relations, which were obtained, were simply incorrect.

\section{Green's function of free ghost field}
\label{sec-ghost}

In this section we write down the Green's function of free ghost field using the identity (\ref{gluon-prop-id}) which holds for every component of the contour function $D$ and $\Delta$.  With the explicit expressions of the gluon functions given by Eqs.~(\ref{D->}, \ref{D-<}, \ref{D-c}, \ref{D-a}) the relation (\ref{gluon-prop-id}) together with (\ref{STI-long-G-1}, \ref{STI-long-G-2}) provides
\ba
\label{G->}
\Delta_{ab}^>(p)&=& - \delta^{ab}  \frac{i\pi}{E_p}
\Big[ \delta(E_p-p_0)\big(n_g({\bf p})+1\big) + \delta(E_p+p_0)n_g(-{\bf p}) \Big], 
\\[2mm]
\label{G-<}
\Delta_{ab}^<(p)&=& - \delta^{ab}  \frac{i\pi}{E_p}
\Big[ \delta(E_p-p_0)n_g({\bf p}) + \delta(E_p+p_0)\big(n_g(-{\bf p})+1\big) \Big], 
\\[2mm]
\label{G-c}
\Delta_{ab}^c(p)&=& \delta^{ab} 
\Big[\frac{1}{p^2+i0^+} - \frac{i\pi}{E_p}
\Big(\delta(p_0-E_p) n_g({\bf p}) + \delta(p_0+E_p) n_g(-{\bf p}) \Big) \Big], 
\\[2mm]
\label{G-a}
\Delta_{ab}^a(p)&=& - \delta^{ab} 
\Big[\frac{1}{p^2-i0^+} 
+ \frac{i\pi}{E_p} \Big(\delta(p_0-E_p) n_g({\bf p}) + \delta(p_0+E_p) n_g(-{\bf p}) \Big)\Big] .
\ea
As seen, the gluon distribution function $n_g({\bf p})$, which describes the {\it physical} gluons, enters the ghost Green's functions.

The relation (\ref{gluon-prop-id}) provides also the retarded $(+)$, advanced $(-)$, and symmetric $({\rm sym})$ ghost Green's functions
\ba
\label{G-pm}
\Delta_{ab}^\pm (p) &=& \frac{\delta_{ab}}{p^2 \pm i \textrm{sgn}(p_0)0^+}, 
\\[2mm]
\label{G-sym}
\Delta_{ab}^{\textrm{sym}}(p)&=& - \delta^{ab} \frac{i\pi}{E_p}
\Big[ \delta(E_p-p_0)\big(2n_g({\bf p})+1\big) + \delta(E_p+p_0)\big(2n_g(-{\bf p})-1\big) \Big], 
\ea
which are used in the subsequent section.

\section{Gluon polarization tensor}
\label{sec-applic}

As an application of the Green's functions of the free ghost field, which are derived in the previous sections, and of the Slavnov-Taylor identity, which requires transversality of the gluon polarization tensor, we discuss here the retarded polarization tensor of a quark-gluon plasma. We note that the Ward-Takahashi identities - abelian analogs of the Slavnov-Taylor identities - were studied in real-time formalism in \cite{Carrington:1998jj}, see also \cite{Carrington:1997sq}.  Our computation is performed within the hard loop approach, see the reviews  \cite{Blaizot:2001nr,Kraemmer:2003gd}, which was generalized to anisotropic systems in \cite{Mrowczynski:2000ed,Mrowczynski:2004kv}. The retarded polarization tensor is an important characteristic of a plasma system, as it carries information about its chromodynamic properties like collective excitations or screening lengths. 

\begin{figure}[t]
\centering
\includegraphics*[width=0.75\textwidth]{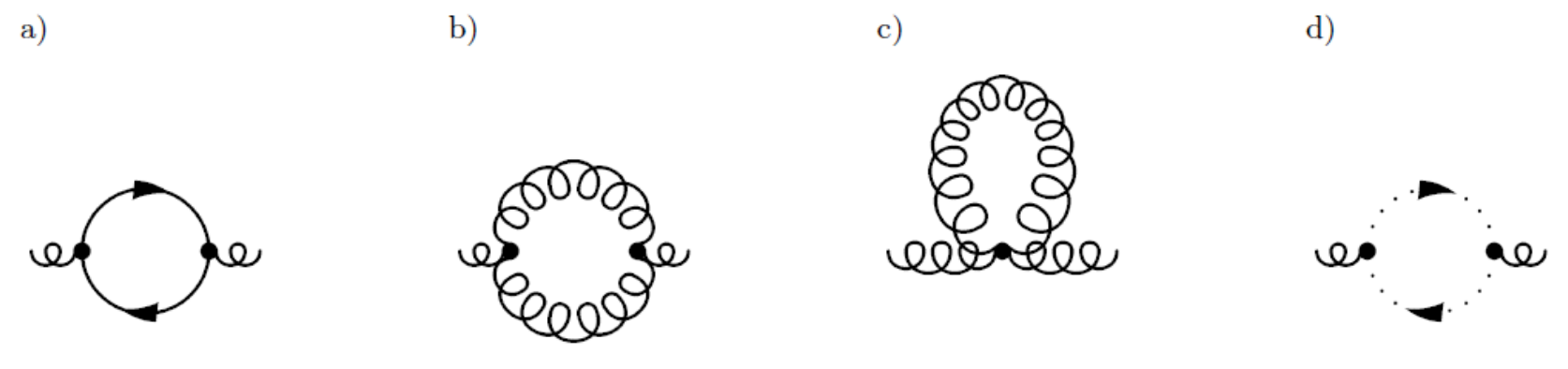}
\caption{The one-loop contributions to the gluon polarization tensor. }
\label{fig-gluon}
\end{figure}

The gluon polarization tensor $\Pi^{\mu \nu}$ can be defined by means of the Dyson-Schwinger equation
\be
i{\cal D}^{\mu \nu} (k) = i D^{\mu \nu} (k)
+ i D^{\mu \rho}(k) \, i\Pi_{\rho \sigma}(k)  \, i{\cal D}^{\sigma \nu}(k) ,
\ee
where ${\cal D}^{\mu \nu}$ and $D^{\mu \nu}$ is the interacting and free gluon propagator, respectively. The lowest order contributions to gluon polarization tensor are given by four diagrams shown in Fig.~\ref{fig-gluon}. The curly, plain and doted lines denote, respectively, gluon, quark and ghost fields.

Applying the Feynman rules, the contribution to the contour $\Pi$ coming from the quark loop corresponding to the graph in  Fig.~\ref{fig-gluon}a  is immediately written down in the coordinate space as
\be
\label{contour-Pi}
_{(a)}\Pi^{\mu \nu}_{ab}(x,y)
= -ig^2 N_c \delta_{ab}
{\rm Tr} [\gamma^\mu S_{ij}(x,y) \gamma^\nu S_{ji}(y,x)] .
\ee
where $S_{ij}(x,y)$ is the quark contour Green's function and the trace is taken over spinor indices. The factor $(-1)$ due to the fermion loop is included and the relation $f_{acd }f_{bcd}=\delta_{ab} N_c$ is used here. 

Since we are interested in the retarded polarization tensor which is expressed through $\Pi^\lg$ as
\be
\label{Pi-pm-Pi><}
\Pi^+ (x,y) =  \Theta(x_0 - y_0)
\Big( \Pi^> (x,y) - \Pi^< (x,y) \Big), 
\ee
the polarization tensors $\Pi^\lg$ are found from the contour tensor (\ref{contour-Pi}) by locating the argument $x_0$ on the upper (lower) and $y_0$ on the lower (upper) branch of the contour. Then, one gets
\be
\label{Pi->-<-x-y}
\big( {_{(a)} \Pi^{\lg} (x,y)} \big)^{\mu \nu}_{ab}
= \frac{i}{2} g^2  \delta^{ab}
{\rm Tr} [\gamma^\mu S_{ij}^\lg (x,y) \gamma^\nu S_{ji}^\gl (y,x)] .
\ee

As the system under study is assumed to be translationally invariant and $S(x,y)= S(x-y)$, we put $y=0$ and write $S(x,y)$ as $S(x)$ and $S(y,x)$ as $S(-x)$. Then, Eq.~(\ref{Pi->-<-x-y}) is
\be
\label{Pi->-<-x-y-1}
\big(  {_{(a)} \Pi^{\lg} (x)} \big)^{\mu \nu}_{ab}
= \frac{i}{2} g^2 \delta^{ab}
{\rm Tr} [\gamma^\mu S_{ij}^\lg (x) \gamma^\nu S_{ji}^\gl (-x)] .
\ee
Since the functions $S^{\pm}$ are expressed through $S^\gl$ analogously to Eq.~(\ref{Pi-pm-Pi><}), the Fourier transformed retarded polarization tensor $\Pi^+ (k)$ is found as
\be
\label{Pi-k-e-1}
\big(  {_{(a)} \Pi^+ (k)} \big)^{\mu \nu}_{ab}
= i\frac{g^2}{4}  \delta^{ab}
 \int \frac{d^4p}{(2\pi )^4}
{\rm Tr} \big[\gamma^\mu S_{ij}^+(p+k)\gamma ^\nu S_{ji}^{\rm sym}(p)
+ \gamma ^\mu S_{ij}^{\rm sym}(p) \gamma ^\nu S_{ji}^-(p-k) \big].
\ee

Further on the index $+$ is dropped and $\Pi^+$ is denoted as  $\Pi$, as only the retarded polarization tensor is discussed. Substituting the functions $S^{\pm}, S^{\rm sym}$ given by Eqs. (\ref{S-pm}, \ref{S-sym}) into the formula (\ref{Pi-k-e-1}), one finds
\ba
\label{Pi-k-e-4}
 _{(a)} \Pi^{\mu \nu}_{ab}(k)
&=& 
- g^2   \delta^{ab} 
\int \frac{d^3p}{(2\pi )^3} \, \frac{n_q ({\bf p}) + \bar{n}_q ({\bf p}) -1}{E_p}
\\ \nonumber
&& \times
\bigg(
\frac{2p^\mu p^\nu  + k^\mu p^\nu + p^\mu k^\nu - g^{\mu \nu} (k \cdot p)  }
{(p+k)^2 + i\, {\rm sgn}\big((p+k)_0\big)0^+}
+ \frac{2p^\mu p^\nu  -  k^\mu p^\nu - p^\mu k^\nu + g^{\mu \nu} (k \cdot p) }
{(p-k)^2 - i\, {\rm sgn}\big((p-k)_0\big)0^+}
\bigg) ,
\ea
where $p^\mu \equiv (E_p, {\bf p})$ with $E_p \equiv |{\bf p}|$,  the traces of gamma matrices are computed and it is taken into account that $p^2 =0$. We also note that after performing the integration over $p_0$, the momentum ${\bf p}$ was changed into  $-{\bf p}$ in the negative energy contribution. 

In the hard loop approximation, when $p \gg k$, we have
\ba
\label{HLA-plus}
\frac{1}{(p+k)^2 + i0^+}
+ \frac{1} {(p-k)^2 - i0^+}
&=& \frac{2k^2}{(k^2)^2 - 4 (k\cdot p)^2 - i {\rm sgn}(k\cdot p) 0^+}
\approx -\frac{1}{2}  \frac{k^2}{(k\cdot p + i 0^+)^2} ,
\\ [2mm]
\label{HLA-minus}
\frac{1}{(p+k)^2 + i0^+}
- \frac{1} {(p-k)^2 - i0^+}
&=& \frac{4(k \cdot p)}{(k^2)^2 - 4 (k\cdot p)^2 - i {\rm sgn}(k\cdot p) 0^+}
\approx \frac{k\cdot p}{(k\cdot p + i 0^+)^2}.
\ea
We note that $(p+k)_0 > 0$ and $(p-k)_0 > 0$ for $p \gg k$. With the formulas (\ref{HLA-plus}, \ref{HLA-minus}), Eq.~(\ref{Pi-k-e-4}) gives
\ba
\label{Pi-k-e-final}
 _{(a)} \Pi^{\mu \nu}_{ab}(k)
 &=&
g^2    \delta^{ab}
 \int \frac{d^3p}{(2\pi )^3} \, \frac{n_q ({\bf p}) + \bar{n}_q ({\bf p}) -1}{E_p} \,
\frac{k^2 p^\mu p^\nu  -  \big(k^\mu p^\nu + p^\mu k^\nu 
- g^{\mu \nu} (k \cdot p) \big) (k \cdot p)}
{(k\cdot p + i 0^+)^2} ,
\ea
which has the well-known structure of the polarization tensor of gauge bosons in ultrarelativistic QED and QCD plasmas.  As seen, the tensor is symmetric with respect to Lorentz indices $ {_{(a)}\Pi}^{\mu \nu}_{ab}(k) =  {_{(a)}\Pi}^{\nu \mu}_{ab}(k)$ and transverse $k_\mu  {_{(a)} \Pi}^{\mu \nu}_{ab}(k) = 0$, as required by the gauge invariance. When $n_q$ and $\bar{n}_q$ both vanish, the polarization tensor (\ref{Pi-k-e-final}) is still nonzero. It is actually infinite and it represents the vacuum effect. Eq.~(\ref{Pi-k-e-final}) gives the contribution of massless quarks  of one flavor. The integral should be multiplied by $N_f$ to get the contribution of $N_f$ flavors of massless quarks.

In analogy to the quark-loop expression (\ref{Pi-k-e-1}), one finds the gluon-loop contribution to the retarded polarization tensor shown in Fig.~\ref{fig-gluon}b as
\ba
\nonumber
 _{(b)}\Pi^{\mu \nu}_{ab}(k) &=& - i\frac{g^2}{4} N_c \delta_{ab}
 \int \frac{d^4p}{(2\pi )^4}  \int \frac{d^4q}{(2\pi )^4} D^{\rm sym}_0(p)
\Big[ (2\pi)^4 \delta^{(4)}(k+p-q)
M^{\mu \nu} (k,q,p) D^+_0(q)
\\ [2mm]
\label{Pi-gluon-loop-2}
&& \;\;\;\;\;\;\;\;\;\;\;\;\;\;\;\;\;\;\;\;\;\;\;\;\;\;\;\;\;\;\;\;\;\;\;\;\;\;\;\;\;\;\;\;\;\;\;\;\;\;\;\;\;\;\;\;\;\;
+ (2\pi)^4 \delta^{(4)}(k-p+q)
M^{\mu \nu} (k,-q,-p) D^-_0(q)
\Big],
\ea
where $D^\pm_0$ and $D^{\rm sym}_0$ are the free gluon Green's functions $D^\pm$ and $D^{\rm sym}$ given by Eqs.~(\ref{D-pm}, \ref{D-sym}) stripped off the Lorentz and color factors that is $D^{\mu \nu}_{ab}(k) = g^{\mu \nu} \delta_{ab} D_0(k)$. The combinatorial factor $1/2$ is included in Eq.~(\ref{Pi-gluon-loop-2}) and 
\be
\label{tensor-M-def}
M^{\mu \nu} (k,q,p) \equiv
\Gamma^{\mu \sigma \rho} (k,-q,p)
\Gamma^{\;\;\,\nu}_{\sigma \;\; \rho} (q,-k,-p) 
\ee
with the three-gluon coupling
\be
\label{3-g-vertex-2}
\Gamma^{\mu \nu \rho} (k,p,q) \equiv
g^{\mu \nu }(k-p)^\rho
+g^{\nu \rho}(p-q)^\mu +g^{\rho \mu}(q-k)^\nu .
\ee

Within the hard loop approximation the tensor (\ref{tensor-M-def}) is computed as
\be
\label{tensor-M-F-HL}
M^{\mu \nu} (k,p \pm k, \pm p) \approx \pm 2 g^{\mu \nu} (k\cdot p)
+ 10 p^\mu p^\nu
\pm 5(k^\mu p^\nu + p^\mu k^\nu),
\ee
where we have taken into account that $p^2=0$. Substituting the expressions (\ref{tensor-M-F-HL}) into Eq.~(\ref{Pi-gluon-loop-2}) and using  the explicit form of the functions $D^\pm$ and $D^{\rm sym}$, we get
\ba
\label{Pi-gluon-loop-5}
 _{(b)}\Pi^{\mu \nu}_{ab}(k) =
\frac{g^2}{4} N_c \delta_{ab}
 \int \frac{d^3p}{(2\pi )^3} \frac{2n_g({\bf p})+1}{E_p}
\frac{5k^2 p^\mu p^\nu - 2 g^{\mu \nu} (k\cdot p)^2
- 5(k^\mu p^\nu + p^\mu k^\nu)(k\cdot p)}{(k\cdot p + i 0^+)^2}.
\ea

The gluon-tadpole contribution to the retarded polarization tensor, which is shown in Fig.~\ref{fig-gluon}c,
equals
\be
\label{Pi-gluon-tadpole-1}
_{(c)}\Pi^{\mu \nu}_{ab}(k) = - i \frac{g^2}{2}
 \int \frac{d^4p}{(2\pi )^4}
\Gamma^{\mu \nu \rho}_{abcc \rho} D^<(p)  ,
\ee
where the combinatorial factor $1/2$ is included and the four-gluon coupling $\Gamma^{\mu \nu \rho \sigma }_{abcd}$ equals
\be
\label{4-g-vertex}
\Gamma^{\mu \nu \rho \sigma }_{abcd} \equiv
f_{abe}f_{ecd}(g^{\mu \sigma} g^{\nu \rho} - g^{\mu \rho} g^{\nu \sigma})
+ f_{ace}f_{edb}(g^{\mu \rho} g^{\nu \sigma} - g^{\mu \nu} g^{\rho \sigma})
+ f_{ade}f_{ebc}(g^{\mu \nu} g^{\rho \sigma} - g^{\mu \sigma} g^{\nu \rho}).
\ee
With the explicit form of the function $D^<(p)$ given by Eq.~(\ref{D-<}), the formula (\ref{Pi-gluon-tadpole-1}) provides
\be
_{(c)}\Pi^{\mu \nu}_{ab}(k) = \frac{3}{2} g^2 N_c \, \delta_{ab} g^{\mu \nu}
\int \frac{d^3p}{(2\pi )^3} \frac{2 n_g({\bf p}) +1}{E_p} .
\ee

The ghost-loop contribution to the retarded polarization tensor, which is shown in Fig.~\ref{fig-gluon}d, equals
\ba
\label{Pi-ghost-loop-2}
 _{(d)}\Pi^{\mu \nu}_{ab}(k) &=&  i\frac{g^2}{2} N_c \delta_{ab}
 \int \frac{d^4p}{(2\pi )^4} \;\Delta^{\rm sym}(p)
\Big[ (p+k)^\mu p^{\nu} \Delta^+(p+k)
+ p^\mu (p-k)^\nu \Delta^-(p-k) \Big] ,
\ea
where the factor $(-1)$  is included as we deal with a fermion loop and the color factor is put in front of the integral. Using the explicit forms of the functions $\Delta^\pm$ and $\Delta^{\rm sym}$ which are given by Eqs.~(\ref{G-pm}, \ref{G-sym}), the formula (\ref{Pi-ghost-loop-2}) is manipulated to
\be
\label{Pi-ghost-loop-4}
 _{(d)}\Pi^{\mu \nu}_{ab}(k) =  -\frac{g^2}{4} N_c \delta_{ab}
\int \frac{d^3p}{(2\pi )^3} \; \frac{2n_g({\bf p})+1}{E_p}
\frac{k^2 p^\mu p^\nu - (k^\mu p^\nu + p^\mu k^\nu) (k\cdot p)}{(k\cdot p + i 0^+)^2},
\ee
which holds in the hard loop approximation.

As already mentioned, the quark-loop contribution to the retarded polarization tensor (\ref{Pi-k-e-final}) is symmetric and transverse with respect to Lorentz indices. The same holds for the sum of the contributions of pure gluodynamics: gluon-loop, gluon-tadpole and ghost-loop. The complete QCD result is obtained by summing up all four contributions and subtracting the vacuum effect. Then, one gets the final formula 
\ba
\label{Pi-final}
 \Pi^{\mu \nu}_{ab}(k) 
&=&
g^2 \delta^{ab}
 \int \frac{d^3p}{(2\pi )^3} 
\frac{n_q({\bf p}) + \bar{n}_q({\bf p}) + 2N_c n_g({\bf p})}{E_p}
\frac{g^{\mu \nu} (k\cdot p)^2
- (k^\mu p^\nu + p^\mu k^\nu) (k\cdot p) + k^2 p^\mu p^\nu}{(k\cdot p + i 0^+)^2},
\ea
which is obviously symmetric and transverse. To our best knowledge this is the first computation of the complete QCD polarization tensor in hard loop approximation performed in the Keldysh-Schwinger (real time) formalism which gives automatically the transversal tensor.  In Refs.~\cite{Weldon:1982aq,Mrowczynski:2000ed}, where the equilibrium and non-equilibrium anisotropic plasmas were considered, respectively, the transversality of $\Pi^{\mu \nu}(k)$ was actually assumed.  In case of  imaginary time formalism, the computation of  the gluon polarization tensor in the hard loop approximation is the textbook material \cite{Kapusta-Gale,lebellac}. We note that the structure of polarization tensor of gauge bosons in hard loop approximation is the same in QED, $\mathcal{N}=1$ SUSY QED \cite{Czajka:2010zh}, QCD and $\mathcal{N}=4$ Super Yang-Mills \cite{Czajka:2012gq}. 

A computation of polarization tensor, which is very similar to that presented above, has been recently done in the context of  $\mathcal{N}=4$ Super Yang-Mills theory in our paper \cite{Czajka:2012gq}. However, the form of free ghost Green's functions (\ref{G->}-\ref{G-sym}) has been postulated with no solid justification. This deficiency has been the motivation of the present study.

\section{Summary, conclusions and outlook}
\label{sec-conclusions}

We have constructed the generating functional of the Keldysh-Schwinger formalism of QCD in a general covariant gauge. The functional provides various relations among the Green's functions, in particular, the perturbative series expressing the interacting Green's functions through the free ones. Deriving the free gluon functions, which are needed for the perturbative calculus, we have found that only the Feynman gauge is free of ill-defined expressions in the Keldysh-Schwinger approach. Using the generating functional, a general Slavnov-Taylor identity has been found. The identity allows one, in particular, to express the ghost Green's function through the gluon one. In this way we managed to obtain the contour Green's function of free ghost field which enters the perturbative series. The functions have been used to compute the retarded gluon polarization tensor in the hard loop approximation. The tensor has appeared to be automatically transverse as required by the gauge symmetry. This opens a possibility to perform other real-time calculations in the Feynman gauge which are usually much simpler than those in physical gauges like the Coulomb one. 

The quark-gluon plasma under consideration has been assumed to be, in general, beyond equilibrium but homogeneous in coordinate space. In other words, the plasma momentum distribution is arbitrary but the system is translationally invariant. The invariance has greatly simplified our analysis but the assumption of homogeneity has to be relaxed to describe a generally non-equilibrium situation. Then, the Fourier transformation is replaced by the Wigner one and one has to refer to the so-called gradient expansion to handle very complex equations. One also faces a difficult problem of interplay of perturbative expansion with the gradient one. These are the problems to be discussed in our subsequent publication.

\section*{Acknowledgments}
We are indebted to A.A. Slavnov for illuminating correspondence. Comments by E. Calzetta and P. Millington are also gratefully acknowledged. This work was partially supported by the Polish National Science Centre under Grant No. 2011/03/B/ST2/00110.


\appendix

\section{More Green's functions of gauge fields}
\label{app-functions-gauge-KS}

Except the functions  ${\cal D}^c$, ${\cal D}^a$, ${\cal D}^>$, ${\cal D}^<$, one often needs the retarded $(+)$, advanced $(-)$ and symmetric $({\rm sym})$ Green's functions which are defined as
\ba
\label{retarded-GF}
i \big({\cal D}_{\mu\nu}^{ab}\big)^{+}(x,y) & \stackrel{{\rm def}}{=} & 
\Theta(x_0-y_0)\frac{{\rm Tr} \big[\rho [A_\mu^a(x), A_\nu^b(y)] \big]}{{\rm Tr}[\rho]}, 
\\[2mm]
\label{advanced-GF}
i \big({\cal D}_{\mu\nu}^{ab}\big)^{-}(x,y) & \stackrel{{\rm def}}{=} & 
- \Theta(y_0-x_0)\frac{{\rm Tr} \big[\rho [A_\mu^a(x), A_\nu^b(y)] \big]}{{\rm Tr}[\rho]}, 
\\[2mm]
\label{symmetric-GF}
i \big({\cal D}_{\mu\nu}^{ab}\big)^{\rm sym}(x,y) & \stackrel{{\rm def}}{=} & 
\frac{{\rm Tr} \big[\rho \{A_\mu^a(x), A_\nu^b(y)\} \big]}{{\rm Tr}[\rho]},
\ea
where $[\ldots,\ldots]$ indicates a commutator and $\{ \ldots,\ldots\}$ an anticommutator of operators. The retarded Green's function ${\cal D}^+$ describes the propagation of both particle and antiparticle disturbance forward in time, while ${\cal D}^-$ governs the evolution backward in time. The functions ${\cal D}^+$, ${\cal D}^-$, ${\cal D}^{\rm sym}$ can be expressed through ${\cal D}^>$, ${\cal D}^<$, ${\cal D}^c$ as
\ba
\label{retarded-GF-id}
{\cal D}^{+}(x,y) & =& \Theta(x_0-y_0) \big({\cal D}^>(x,y) - {\cal D}^<(x,y) \big) = {\cal D}^c(x,y) -  {\cal D}^<(x,y) , 
\\[2mm]
\label{advanced-GF-id}
{\cal D}^-(x,y) &=& \Theta(y_0-x_0) \big( {\cal D}^<(x,y) - {\cal D}^>(x,y)\big) = {\cal D}^c(x,y) -  {\cal D}^>(x,y), 
\\[2mm]
\label{symmetric-GF-id}
{\cal D}^{\rm sym}(x,y) & = & {\cal D}^>(x,y) + {\cal D}^<(x,y).
\ea
Using the relations (\ref{retarded-GF-id}, \ref{advanced-GF-id}, \ref{symmetric-GF-id}) together with the functions $D^>$, $D^<$, $D^c$  derived in Sec.~\ref{sec-free-funcs}, one easily obtains the retarded, advanced and symmetric Green's functions of {\it free} fields as
\ba
\label{D-pm}
\big(D^{ab}_{\mu\nu}\big)^\pm (p) &=&- \frac{g_{\mu\nu} \delta^{ab}}{p^2 \pm i\textrm{sgn}(p_0)0^+},
\\ [2mm]
\label{D-sym}
\big(D^{ab}_{\mu\nu}\big)^{\textrm{sym}}(p) &=& g_{\mu\nu} \delta^{ab} \frac{i\pi}{E_p}
\Big[ \delta(E_p-p_0) \big(2n_g({\bf p}) +1\big) +\delta(E_p+p_0) \big(2n_g(-{\bf p})+1\big) \Big].
\ea

\section{Green's functions of fermion field}
\label{app-functions-quark-KS}

The Green's functions of a fermion field are defined analogously to those of vector one, see {\it e.g.} \cite{Mrowczynski:1992hq}, and a technique to derive the free functions is also similar. Therefore, we only list here some formulas we need for the calculations presented in Sec.~\ref{sec-applic}. The Green's functions of the free massless quark field are
\ba
\label{S-pm}
S^{\pm}_{ij}(p) &=& \frac{\delta_{ij} {p\sla}}{p^2\pm i\, {\rm sgn}(p_0)0^+},
\\
\label{S->}
S^>_{ij}(p) &=&  \delta_{ij} \frac{i\pi}{E_p} {p\sla}
\Big( \delta (E_p - p_0)  \big[ n_q ({\bf p}) -1\big]
+ \delta (E_p + p_0) \bar{n}_q (-{\bf p}) \Big),
\\
\label{S-<}
S^<_{ij}(p) &=& \delta_{ij} \frac{i\pi}{E_p} {p\sla} \Big( \delta (E_p - p_0)  n_q({\bf p})
+ \delta (E_p + p_0) \big[\bar{n}_q (-{\bf p}) - 1\big] \Big),
\\
\label{S-sym}
S^{\rm sym}_{ij}(p) & =& \delta_{ij} \frac{i\pi}{E_p} {p\sla}
\Big( \delta (E_p - p_0)   \big[ 2n_q({\bf p}) -1 \big]
+ \delta (E_p + p_0) \big[ 2 \bar{n}_q (-{\bf p}) - 1\big] \Big),
\ea
where $i,j = 1,\,2, \dots N_c$ are color indices of the fundamental representation, $n_q({\bf p})$ and $\bar{n}_q({\bf p})$ are the distribution functions of quarks and antiquarks, respectively, which are assumed to be unpolarized with respect to spin and color degrees of freedom. The distribution function is normalized in such a way that the quark density of a given flavor equals
\be
\rho_q = 2 N_c \int \frac{d^3p}{(2\pi)^3}\, n_q ({\bf p}) ,
\ee
where the factor of 2 takes into account two spin states of each quark. One checks that the functions (\ref{S-pm}, \ref{S->}, \ref{S-<}) obey the identity $S^>(p) - S^< (p) = S^+(p) - S^-(p)$.


\end{document}